\documentclass[conference]{IEEEtran}
\ifCLASSINFOpdf
\else
\fi
\usepackage{booktabs} 
\usepackage{graphicx}
\usepackage{subfig}
\usepackage[dvipsnames]{xcolor}

\begin{document}
%
\title{Audio-Visual Evaluation of Oratory Skills}

\author{\IEEEauthorblockN{Tzvi Michelson ~~~~~~~~~~~~~ Shmuel Peleg}
\IEEEauthorblockA{School of Computer Science and Engineering\\
Hebrew University of Jerusalem\\
Jerusalem, Israel}
}



\maketitle

\begin{abstract}
What makes a talk successful? Is it the content or the presentation? We try to estimate the contribution of the speaker's oratory skills to the talk's success, while ignoring the content of the talk. By oratory skills we refer to facial expressions, motions and gestures, as well as the vocal features. We use TED Talks as our dataset, and measure the success of each talk by its view count. Using this dataset we train a neural network to assess the oratory skills in a talk through three factors: body pose, facial expressions, and acoustic features. 

Most previous work on automatic evaluation of oratory skills uses hand-crafted expert annotations for both the quality of the talk and for the identification of predefined actions. Unlike prior art, we measure the quality to be equivalent to the view count of the talk as counted by TED, and allow the network to automatically learn the actions, expressions, and sounds that are relevant to the success of a talk. We find that oratory skills alone contribute substantially to the chances of a talk being successful.
\end{abstract}


%
\IEEEpeerreviewmaketitle

\begin{figure*}[!t]
\centering
\includegraphics[width=7.3in]{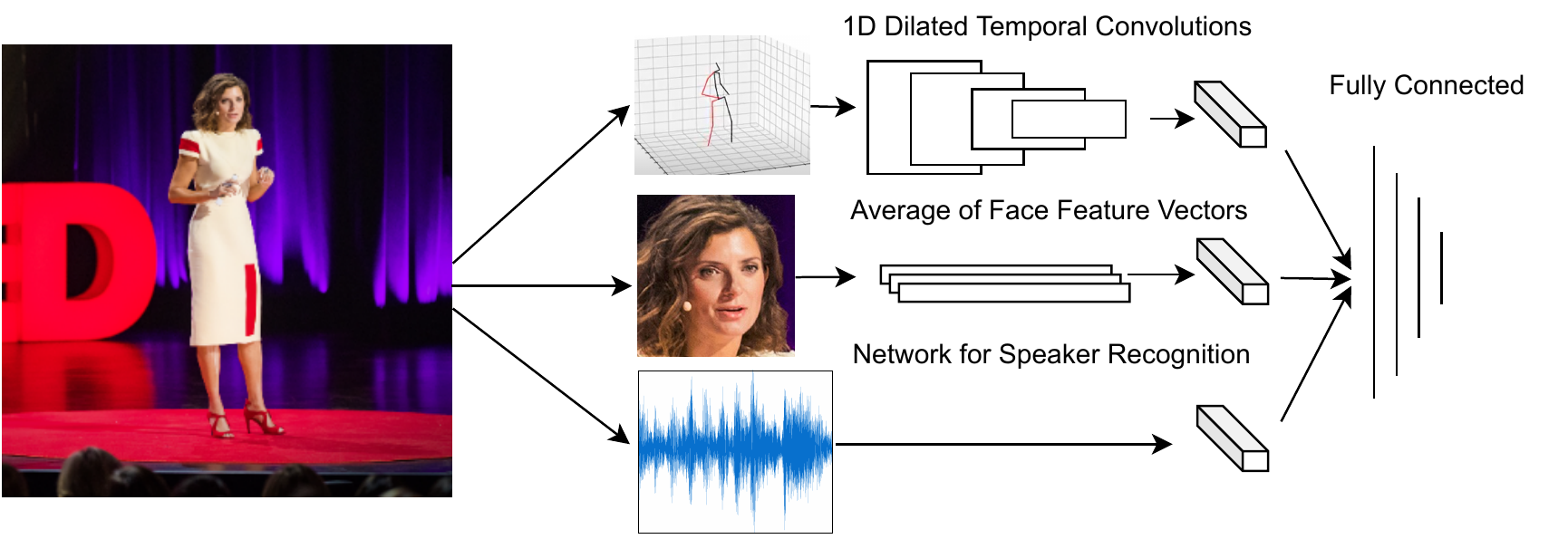}
\hfil
\caption{Our proposed neural network to evaluate oratory skills. The network considers features vectors from three modalities over a sequence of video frames: Pose keypoints, face descriptors, and voice feature vectors. Each feature vector is computed using existing pretrained networks. (i) 3D coordinates of keypoints are convolved using dilated convolutions to produce a single feature vector.  (ii) Face descriptors from consecutive frames in every segment are averaged, and (iii) a single feature vector is extracted from the voice. All three descriptors are concatenated and pass through a few fully connected layers.}
\label{fig_sim}
\end{figure*}

\section{Introduction}
People intuitively know whether a speech is well presented. In addition to presenting meaningful content, aspects such as tone, body posture, hand gestures and facial expressions are important to drive home the message \cite{hoogterp2014your}. In this work we train a model to measure the impact of the oratory skills on the perceived quality of the speech. We utilize publicly available data, namely TED Talks, where the corresponding amount of views at the talk serves as our indicator for the talk's success. Using both video and audio as input, we train a neural network to analyze and score the content-independent oratory skills demonstrated in each TED Talk.

We believe this study is interesting for these reasons:
\begin{enumerate}
   \setlength{\itemsep}{1pt}
   \setlength{\parskip}{1pt}
   \setlength{\parsep}{1pt}
    \item An algorithm that evaluates a public speaker allows him to practice and improve without human aid. The speaker can try different variations and see how the algorithm evaluates each one. We will note that the motivational speakers' market in the U.S. generated an estimated 1.9 billion dollars in revenues in 2019 \cite{U.S.MotivationalSpeaking2020}. 
    \item Public speaking is a field studied in psychology. Today, when researchers want to measure how an intervention impacts a public speaker, they must recruit experts to manually evaluate the quality of the speech before and after the intervention \cite{tanveer2015rhema, schneider2015presentation}. An automatic tool for the analysis of oratory skills will remove this need and significantly ease the research.
    \item Training a model to evaluate oratory skills is part of the ongoing effort of the AI community to teach computers about humans.
\end{enumerate}
\vspace{1pt}

One of the major problems in machine learning is obtaining high quality labelled data. When assessing the problem of image classification, labels can be obtained and verified easily. But many labelling problems are much harder. In cases such as medical diagnosis and facial emotion recognition, a team is deployed to evaluate each image, and individual evaluations often diverge. In addition, even when all evaluations are identical they may still be incorrect. By using TED Talks as our data, with the view count as labels, we circumvent subjective labeling and ensure the quality of our ground truth. We encounter several obstacles when fitting this data to our problem, an explanation of how we address them can be found in Sec.~\ref{sec:dataset}.

Our work contains three major contributions to previous works:
\begin{itemize}
    \item We do not rely on human annotations or expert opinion at any stage. Thus, we ensure our labels are unbiased and detect the effect of subtleties a human expert might miss.
    \item Our training data consists of 290 hours of public speaking whereas previous datasets have contained up to 10 hours.
    \item Most previous work use depth sensors to capture 3D pose, while our approach uses only a video recording (images and sound).
\end{itemize}

\section{Related Work}
Previous work in the field of automatic tools for measurement and improvement of public speaking skills can be split into three main categories.

\subsection{Methods for increasing awareness}

Tanveer et al. presented AutoManner \cite{tanveer2016automanner}, an unsupervised method to automatically extract human gestures using Microsoft Kinect and use this data to make speakers aware of their mannerisms. The Kinect contains three components that work together to detect motion and create a physical image: an RGB  video camera, a depth sensor, and a multi-array microphone. The physical image is processed to recognize repeating mannerisms, which are then displayed to the speaker. This work follows a previous work of same authors \cite{tanveer2015rhema}, which compares the effect of presenting the speaker with an analysis of the acoustic features of his voice vs. giving him an evaluation of his performance. The feedback was given to the speaker via a pair of Google glasses worn while speaking.

\subsection{Analysis from predefined features}

Other papers using Microsoft Kinect to automatically analyze speeches, base their analysis on hand crafted features that experts deem necessary for successful public speaking \cite{dermody2016multimodal, schneider2015presentation, damian2015augmenting, barmaki2015providing} . These features include: eye contact, open body pose, body energy, speech rate, speech volume and more. The RAP system \cite{ochoa2018rap} uses an 8MP Raspicamera v2.1 and a microphone to gather video and audio input. They rate the speech on 4 factors: body posture, gaze, filled pauses (e.g. \textit{hmmmm}s \cite{WikipediaFillerLinguistics}) and audio volume. As before, these factors are predefined by experts as important for the evaluation of speeches.

\subsection{Machine learning prediction of pre-defined features}
A public speaking course given at Eindhoven University of Technology was utilized by Nguyen et al. \cite{nguyen2012online} to video the speakers and gather the feedback they were given. Based on the feedback they trained a k-nearest neighbors classifier on the speaker's postures and gestures as observed by Microsoft Kinect. Their dataset was composed of 32 university employees each one speaking for approximately one minute, with the postures and gestures tagged manually. The classifier predicted whether the input gathered from the sensors fit a predefined posture or gesture.

Gan et al. \cite{gan2015multi} gathered 51 speakers, and had each one speak for approximately 15 minutes totaling 10 hours of recorded speech. Their equipment included a Microsoft Kinect, 2 RGB cameras and three Google Glasses. One of the Google Glasses was worn by the speaker while the other two were worn by randomly selected audience members and used to measure audience attention. The ground truth was annotated manually based on expert opinions. They gathered both visual and acoustic information and used a Support-vector machine classifier with a polynomial kernel for classification.






\section{The TED Talks dataset}
\label{sec:dataset}
We download our data from the TED website \cite{TEDWebsite} using the metadata published on Kaggle \cite{Banik2017TEDTalksKaggle}. We included all single speaker TED Talks published before 21 September 2017. This included 2401 TED Talks totaling 290 hours of speech. The success of each talk was measured by the view count of this talk until Sept. 21, 2017. The view count has a range from 150 thousand to 47 million, with a median of 1.1 million and a mean of 1.7 million.

The frame rate of all videos is set to 24 fps with quality of $640 \times 480$. All shots were broken to  5 second segments. Each segment is analyzed to ensure it only has a single face in every frame, and every segment containing a frame with more than one face is discarded. The analysis is done using the model proposed by Zhang et al. \cite{zhang2016joint}. This procedure ensures no clips of the audience are included. The audio is cleaned using a pretrained model proposed by Westhausen et al. \cite{Westhausen2020}, removing all sounds aside from human speech.

We will note that in every TED Talk there are shots from multiple angles. We do not augment the data to compensate for this, but rather rely on our network to learn appropriate representations.

We assume that the view count of a TED Talk is a noisy function of two dominant parameters: the oratory skills of the speaker and the content of his speech. Our model only attempts to measure the oratory skills. These are usually measured by audience engagement \cite{schreiber2012development}, view count is a very good proxy for this since it is a count of how many people were engaged \cite{rowe2011forecasting, thurman2014newspaper, thelwall2017researchgate}.

Since we used TED talks that were published over 11 years (2006-2017) we calculate the 33rd percentile and 66th percentile of the view counts in every year. All talks which have a lower view count then the 33rd percentile are classified as ``bad" talks (0), and all talks having a larger view count then the 66th percentile are classified as ``good" talks (1). Talks in the middle third are ignored. We address our talk classification problem as the binary classification problem of determining whether a specific TED Talk is a good talk or a bad talk.

\vspace{0.2cm}
\section{Audio-Visual assessment of oratory skills}
We use existing pretrained models to preprocess three modalities:

\subsection{Pose}
Every 5-second video segment is given to a 2D skeleton extractor, and the extracted 2D key points are given to a 3D pose estimator. We utilize a publicly available model from Wu et al. \cite{wu2019detectron2} to extract the 2D key points. Their model returns for every frame two coordinates for each of 17 key points: eyes, ears, nose, shoulders, elbows, wrists, hips, knees and ankles. At the second stage we use the network described by Pavllo et al. \cite{pavllo20193d} to discover the 3D key points. Their model uses the temporal knowledge gained from consecutive frames to map the 2D skeleton to 3D coordinates. The model outputs 3 coordinates for every one of the 17 key points. In our experiments the model worked very well and presented a realistic 3D skeleton. At the end of this step we have for each frame a vector with $17 \times 3 = 51$ entries. Each segment consists of 120 frames, so we end up with a $120 \times 51$ matrix for every segment.

We use 1D dilated convolutions on the temporal axis of the pose estimations to extract a single vector representing the pose and its variations in a each temporal segment. The convolutional network consists of 5 convolutional layers with the following dilation rates: 1, 2, 2, 4, 4 respectively. Every layer is followed by a ReLU, has 32 filters, and has a 5-dimensional kernel. The output of the final layer is flattened, resulting in a single 1088-dimensional vector representing the body pose in the segment. 

\subsection{Face}
The face detection model proposed by Zhang et al. \cite{zhang2016joint} was used to identify the single face that appears in every frame. All video segments containing a frame with more then one face were discarded to avoid catching the audience's reaction. We then use the face recognition model proposed by Cao et al. \cite{cao2018vggface2} to extract a 512-dimensional feature vector representing the face in every frame. Other models did not lead to improved performance. While the model was pretrained for facial recognition, it has been shown that feature vectors extracted from facial recognition models are useful for other tasks that require face representation \cite{ephrat2018looking}. Due to computational limitations we cannot analyze every feature vector for every frame in every segment, so we average all feature vectors in the 5 second video segment, giving a single 512-dimensional feature vector representing the facial expressions in a single video segment.

\subsection{Voice}
The speaker recognition network proposed by Desplanques et al. \cite{DBLP:conf/interspeech/DesplanquesTD20}, made available by \cite{speechbrain}, was used to extract a 512-dimensional vector representing the speech in the audio of every video segment. Here as well other tested models did not improve performance.

\subsection{Implementation details}
The three vectors (pose, face, and voice) are concatenated, to form a 2112-dimensional vector which is normalized using Batch Normalization.
The network continues with four fully connected layers with ReLUs and Batch Normalizations interwoven. The layer's dimensions are 2048, 1024, 512, 256 respectively. The network finishes with a sigmoid, and uses binary cross entropy as its loss function with the Adam optimizer \cite{kingma2014adam} and an initial learning rate of $5 \times 10^{-6}$. A visualization of the network can be found in Figure~\ref{fig_sim}.


\begin{table}[tb]
\centering
\small

\begin{tabular}{lccc}
\toprule
\textbf{Inputs} & \textbf{Model} & \textbf{ROC AUC} & \textbf{F1 Score} \\
\midrule
Video & RAP & 0.53 & 0.54\\
Audio & RAP & 0.52 & 0.53\\
Video + Audio & RAP & 0.55 & 0.56 \\
\midrule
Face & Ours & 0.58 & 0.59 \\
Voice & Ours & 0.60 & 0.61 \\
Pose & Ours & 0.57 & 0.57 \\
\midrule
Face + Voice & Ours & 0.63 & 0.62 \\
Pose + Voice & Ours & 0.63 & 0.63 \\
Face + Pose & Ours & 0.61 & 0.61 \\
\midrule
Face + Pose + Voice (mean) & Ours & 0.63 & 0.63 \\
Face + Pose + Voice (median) & Ours & 0.63 & 0.64 \\
Face + Pose + Voice (max) & Ours & \textbf{0.65} & \textbf{0.67} \\
\bottomrule
\end{tabular}
\vspace{0.25cm}
\caption{Effects of different input modalities vs. prior work.\\
ROC AUC and F1 score regard the classification of a TED Talk as good or bad by the model vs. the view count. The scores are averaged over 5-fold cross-validation.\\
Mean, median, and max are different combinations of the results for the multiple video segments of the same TED Talk.}
\label{tab:main_method_results}
\end{table}

\section{Experiments}

Our network gives a score between zero and one for every 5-second video segment, and each TED talk has multiple such segments. We need combine all segment scores to a single score. Using the maximum of all segment scores gave best results. We hypothesize that in most talks there are specific points in time when a speaker succeeds in capturing his audience. These points discriminate between successful and unsuccessful speakers. Successful speakers will have more of these points and they will be of higher quality. When comparing maximums we are in essence comparing two of these points from two different talks. Using the maximum score gave a small improvement upon using the mean or the median of the segment scores.

We test our method on various parts of our input. In every test 5-fold cross-validation is used to ensure the consistency of our results. We also compare our method to RAP \cite{ochoa2018rap}, which does not depend on a kinect sensor. RAP uses body posture, similar to our pose estimation, and uses filled pauses (\textit{hmmmm}s \cite{WikipediaFillerLinguistics}) as well as audio volume to judge the speaking quality. Since the RAP software is not publicly available, we had to recreate it based on the description in \cite{ochoa2018rap}. To make the RAP system applicable to our dataset we took the outputs of their system, fed them to a classifier and trained on our data. We experimented with a number of classifiers, the results can be found in Table ~\ref{tab:main_method_results}.

Interestingly, we find that when replacing the latent vector representing the voice or face in a low-scoring segment  with a high-scoring latent vector representing the same modality, the segment score improves significantly. Replacing the pose latent vector leads to an improvement 70\% of the time, but the improvement is minor. This fact, independently, is interesting for public speakers, yet can also be used to provide modality feedback from our model. It is possible to iteratively replace each modality and see which replacement leads to the greatest improvement, that is the modality the speaker must improve.

\section{Conclusion}
We proposed an audio-visual neural network based model for end-to-end evaluation of oratory skills. The model uses three features: body pose, facial attributes and acoustic features. We succeed in reaching ROC AUC of 0.65 with F1 Score of 0.67 on a noisy, organic, in the wild dataset and improve on previous work in the field. Despite our success we believe there is significant room for improvement. Further research could attempt to model the very short term dependencies between different modalities to improve on our results. Furthermore, we believe that unsupervised pretraining specialized to create good representations of human behaviour could also improve results significantly although the appropriate task has yet to be suggested. 

\section{Acknowledgments}
This research was supported by grants from the Israel Science Foundation, the DFG, and the Crown Family Foundation.






%
{
\bibliographystyle{IEEEtran}
\bibliography{refs}

\begin{thebibliography}{10}
\providecommand{\url}[1]{#1}
\csname url@samestyle\endcsname
\providecommand{\newblock}{\relax}
\providecommand{\bibinfo}[2]{#2}
\providecommand{\BIBentrySTDinterwordspacing}{\spaceskip=0pt\relax}
\providecommand{\BIBentryALTinterwordstretchfactor}{4}
\providecommand{\BIBentryALTinterwordspacing}{\spaceskip=\fontdimen2\font plus
\BIBentryALTinterwordstretchfactor\fontdimen3\font minus
  \fontdimen4\font\relax}
\providecommand{\BIBforeignlanguage}[2]{{%
\expandafter\ifx\csname l@#1\endcsname\relax
\typeout{** WARNING: IEEEtran.bst: No hyphenation pattern has been}%
\typeout{** loaded for the language `#1'. Using the pattern for}%
\typeout{** the default language instead.}%
\else
\language=\csname l@#1\endcsname
\fi
#2}}
\providecommand{\BIBdecl}{\relax}
\BIBdecl

\bibitem{hoogterp2014your}
B.~Hoogterp, \emph{Your Perfect Presentation: Speak in Front of Any Audience
  Anytime Anywhere and Never Be Nervous Again}.\hskip 1em plus 0.5em minus
  0.4em\relax McGraw Hill Professional, 2014.

\bibitem{U.S.MotivationalSpeaking2020}
J.~LaRosa, ``U.s. motivational speaking industry worth 1.9 billion,''
  \url{https://blog.marketresearch.com/u.s.-motivational-speaking-market-worth-1.9-billion},
  2020.

\bibitem{tanveer2015rhema}
M.~I. Tanveer, E.~Lin, and M.~Hoque, ``Rhema: A real-time in-situ intelligent
  interface to help people with public speaking,'' in \emph{Proceedings of the
  20th International Conference on Intelligent User Interfaces}, 2015, pp.
  286--295.

\bibitem{schneider2015presentation}
J.~Schneider, D.~B{\"o}rner, P.~Van~Rosmalen, and M.~Specht, ``Presentation
  trainer, your public speaking multimodal coach,'' in \emph{Proceedings of the
  2015 ACM on International Conference on Multimodal Interaction}, 2015, pp.
  539--546.

\bibitem{tanveer2016automanner}
M.~I. Tanveer, R.~Zhao, K.~Chen, Z.~Tiet, and M.~E. Hoque, ``Automanner: An
  automated interface for making public speakers aware of their mannerisms,''
  in \emph{Proceedings of the 21st International Conference on Intelligent User
  Interfaces}, 2016, pp. 385--396.

\bibitem{dermody2016multimodal}
F.~Dermody and A.~Sutherland, ``Multimodal system for public speaking with real
  time feedback: a positive computing perspective,'' in \emph{Proceedings of
  the 18th ACM International Conference on Multimodal Interaction}, 2016, pp.
  408--409.

\bibitem{damian2015augmenting}
I.~Damian, C.~S. Tan, T.~Baur, J.~Sch{\"o}ning, K.~Luyten, and E.~Andr{\'e},
  ``Augmenting social interactions: Realtime behavioural feedback using social
  signal processing techniques,'' in \emph{Proceedings of the 33rd annual ACM
  conference on Human factors in computing systems}, 2015, pp. 565--574.

\bibitem{barmaki2015providing}
R.~Barmaki and C.~E. Hughes, ``Providing real-time feedback for student
  teachers in a virtual rehearsal environment,'' in \emph{Proceedings of the
  2015 ACM on International Conference on Multimodal Interaction}, 2015, pp.
  531--537.

\bibitem{ochoa2018rap}
X.~Ochoa, F.~Dom{\'\i}nguez, B.~Guam{\'a}n, R.~Maya, G.~Falcones, and
  J.~Castells, ``The rap system: Automatic feedback of oral presentation skills
  using multimodal analysis and low-cost sensors,'' in \emph{Proceedings of the
  8th international conference on learning analytics and knowledge}, 2018, pp.
  360--364.

\bibitem{WikipediaFillerLinguistics}
``Wikipedia - filler (linguistics),''
  \url{\\https://en.wikipedia.org/wiki/Filler\_(linguistics)}, 2021.

\bibitem{nguyen2012online}
A.-T. Nguyen, W.~Chen, and M.~Rauterberg, ``Online feedback system for public
  speakers,'' in \emph{2012 IEEE Symposium on E-Learning, E-Management and
  E-Services}.\hskip 1em plus 0.5em minus 0.4em\relax IEEE, 2012, pp. 1--5.

\bibitem{gan2015multi}
T.~Gan, Y.~Wong, B.~Mandal, V.~Chandrasekhar, and M.~S. Kankanhalli,
  ``Multi-sensor self-quantification of presentations,'' in \emph{Proceedings
  of the 23rd ACM international conference on Multimedia}, 2015, pp. 601--610.

\bibitem{TEDWebsite}
``Ted - ideas worth spreading,'' \url{\\https://www.ted.com/talks}, 2021.

\bibitem{Banik2017TEDTalksKaggle}
R.~Banik, ``Ted talks dataset - kaggle,''
  \url{https://www.kaggle.com/rounakbanik/ted-talks}, 2017.

\bibitem{zhang2016joint}
K.~Zhang, Z.~Zhang, Z.~Li, and Y.~Qiao, ``Joint face detection and alignment
  using multitask cascaded convolutional networks,'' \emph{IEEE Signal
  Processing Letters}, vol.~23, no.~10, pp. 1499--1503, 2016.

\bibitem{Westhausen2020}
\BIBentryALTinterwordspacing
N.~L. Westhausen and B.~T. Meyer, ``{Dual-Signal Transformation LSTM Network
  for Real-Time Noise Suppression},'' in \emph{Proc. Interspeech 2020}, 2020,
  pp. 2477--2481. [Online]. Available:
  \url{http://dx.doi.org/10.21437/Interspeech.2020-2631}
\BIBentrySTDinterwordspacing

\bibitem{schreiber2012development}
L.~M. Schreiber, G.~D. Paul, and L.~R. Shibley, ``The development and test of
  the public speaking competence rubric,'' \emph{Communication Education},
  vol.~61, no.~3, pp. 205--233, 2012.

\bibitem{rowe2011forecasting}
M.~Rowe, ``Forecasting audience increase on youtube,'' 2011.

\bibitem{thurman2014newspaper}
N.~Thurman, ``Newspaper consumption in the digital age: Measuring multi-channel
  audience attention and brand popularity,'' \emph{Digital Journalism}, vol.~2,
  no.~2, pp. 156--178, 2014.

\bibitem{thelwall2017researchgate}
M.~Thelwall and K.~Kousha, ``Researchgate articles: Age, discipline, audience
  size, and impact,'' \emph{Journal of the Association for information Science
  and technology}, vol.~68, no.~2, pp. 468--479, 2017.

\bibitem{wu2019detectron2}
Y.~Wu, A.~Kirillov, F.~Massa, W.-Y. Lo, and R.~Girshick, ``Detectron2,''
  \url{https://github.com/facebookresearch/detectron2}, 2019.

\bibitem{pavllo20193d}
D.~Pavllo, C.~Feichtenhofer, D.~Grangier, and M.~Auli, ``3d human pose
  estimation in video with temporal convolutions and semi-supervised
  training,'' in \emph{Proceedings of the IEEE/CVF Conference on Computer
  Vision and Pattern Recognition}, 2019, pp. 7753--7762.

\bibitem{cao2018vggface2}
Q.~Cao, L.~Shen, W.~Xie, O.~M. Parkhi, and A.~Zisserman, ``Vggface2: A dataset
  for recognising faces across pose and age,'' in \emph{2018 13th IEEE
  international conference on automatic face \& gesture recognition (FG
  2018)}.\hskip 1em plus 0.5em minus 0.4em\relax IEEE, 2018, pp. 67--74.

\bibitem{ephrat2018looking}
A.~Ephrat, I.~Mosseri, O.~Lang, T.~Dekel, K.~Wilson, A.~Hassidim, W.~T.
  Freeman, and M.~Rubinstein, ``Looking to listen at the cocktail party: A
  speaker-independent audio-visual model for speech separation,'' \emph{arXiv
  preprint arXiv:1804.03619}, 2018.

\bibitem{DBLP:conf/interspeech/DesplanquesTD20}
B.~Desplanques, J.~Thienpondt, and K.~Demuynck, ``{ECAPA-TDNN:} emphasized
  channel attention, propagation and aggregation in {TDNN} based speaker
  verification,'' in \emph{Interspeech 2020}, H.~Meng, B.~Xu, and T.~F. Zheng,
  Eds.\hskip 1em plus 0.5em minus 0.4em\relax {ISCA}, 2020, pp. 3830--3834.

\bibitem{speechbrain}
M.~Ravanelli, T.~Parcollet, P.~Plantinga, A.~Rouhe, S.~Cornell, L.~Lugosch,
  C.~Subakan, N.~Dawalatabad, A.~Heba, J.~Zhong, J.-C. Chou, S.-L. Yeh, S.-W.
  Fu, C.-F. Liao, E.~Rastorgueva, F.~Grondin, W.~Aris, H.~Na, Y.~Gao, R.~D.
  Mori, and Y.~Bengio, ``{SpeechBrain}: A general-purpose speech toolkit,''
  2021, arXiv:2106.04624.

\bibitem{kingma2014adam}
D.~P. Kingma and J.~Ba, ``Adam: A method for stochastic optimization,''
  \emph{arXiv preprint arXiv:1412.6980}, 2014.

\end{thebibliography}
}




\end{document}